\documentclass[a4paper,11pt,nobalancelastpage,superscriptaddress,tightenlines]{revtex4-2}
\UseRawInputEncoding
\usepackage{color,ifthen,amsthm,amsmath,amsxtra,amsfonts,dsfont,graphicx,bm,tikz,scalerel,wasysym,bbm,graphicx,amsthm,braket,physics,empheq}
\usepackage[colorlinks=true,linkcolor=blue, citecolor=blue, urlcolor=blue, bookmarks]{hyperref}

\usepackage[T1]{fontenc}
\usepackage[utf8]{inputenc}

\usepackage{amsmath,braket,amssymb}
\usepackage{multirow}
\usepackage{empheq}
\usepackage{tensor}
\usepackage{cancel}
\usepackage{enumitem}
\usepackage{simplewick}

\usepackage{enumitem}
\usepackage{slashed}
\usepackage{graphicx}
\usepackage{xcolor}

\colorlet{Green}{black!30!green}

\definecolor{THc}{rgb}{0.9,0.3,0.2}

\usepackage{tikz}
\usetikzlibrary{calc,fadings,decorations.pathreplacing,shapes,shapes.multipart,arrows,shapes.misc,intersections,positioning,patterns}
\tikzset{arrow data/.style 2 args={%
		decoration={%
			markings,
			mark=at position #1 with \arrow{#2}},
		postaction=decorate}
}
\usepackage{bm}
\usepackage[colorlinks=true,citecolor=blue,linkcolor=blue,urlcolor=blue]{hyperref}
\usepackage{dsfont}
\usepackage{changepage}
\usepackage{array}
\usepackage{soul}
\usepackage[capitalise]{cleveref}
\crefname{section}{Sec.}{Secs.}
\Crefname{section}{Sec.}{Secs.}
\usepackage{xifthen}
\usepackage{xargs}
\usepackage{hhline}
\usepackage{pifont}

\usepackage{amsthm}
\theoremstyle{definition}

\theoremstyle{plain}

\usepackage{bm}

\graphicspath{{./}{./figures/}}

\newcommand{\bit}{\begin{itemize}}
	\newcommand{\eit}{\end{itemize}}

\renewcommand{\>}{\right\rangle}
\newcommand{\<}{\left\langle}
\newcommand{\ba}{\begin{align}}
	\newcommand{\ea}{\end{align}}
\newcommand{\be}{\begin{equation}}
	\newcommand{\ee}{\end{equation}}
\newcommand{\bi}{\begin{itemize}}
	\newcommand{\ei}{\end{itemize}}

\DeclareMathAlphabet{\mymathbb}{U}{BOONDOX-ds}{m}{n}

\usepackage{physics}

\usepackage{braket}
\usepackage[normalem]{ulem}

\begin{document}
	\date{\today}

	\newcommand{\bbra}[1]{\<\< #1 \right|\right.}
	\newcommand{\kket}[1]{\left.\left| #1 \>\>}
	\newcommand{\bbrakket}[1]{\< \Braket{#1} \>}
	\newcommand{\pll}{\parallel}
	\newcommand{\nn}{\nonumber}
	\newcommand{\transp}{\text{transp.}}
	\newcommand{\nor}{z_{J,H}}
	
	\newcommand{\hL}{\hat{L}}
	\newcommand{\hR}{\hat{R}}
	\newcommand{\hQ}{\hat{Q}}

    	\newcommand{\hh}{\mathtt{h}}

	\title{Construction and simulability of quantum circuits with free fermions in disguise}

\begin{abstract}
We provide a systematic construction for local quantum circuits hosting free fermions in disguise, both with staircase and brickwork architectures. Similar to the original Hamiltonian model introduced by Fendley, these circuits are defined by the fact that the Floquet operator corresponding to a single time step can not be diagonalized by means of any Jordan-Wigner transformation, but still displays a free-fermionic spectrum. Our construction makes use of suitable non-local transfer matrices commuting with the Floquet operator, allowing us to establish the free fermionic spectrum. We also study the dynamics of these circuits after they are initialized in arbitrary product states, proving that the evolution of certain local observables can be simulated efficiently on classical computers. Our work proves recent conjectures in the literature and raises new questions on the classical simulability of free fermions in disguise.
\end{abstract}

\author{D\'avid Sz\'asz-Schagrin}
\affiliation{Dipartimento di Fisica e Astronomia, Universit\`a di Bologna and INFN, Sezione di Bologna, via Irnerio 46, I-40126 Bologna, Italy}
\author{Daniele Cristani}
\affiliation{Dipartimento di Fisica e Astronomia, Universit\`a di Bologna and INFN, Sezione di Bologna, via Irnerio 46, I-40126 Bologna, Italy}
\author{Lorenzo Piroli}
\affiliation{Dipartimento di Fisica e Astronomia, Universit\`a di Bologna and INFN, Sezione di Bologna, via Irnerio 46, I-40126 Bologna, Italy}
\author{Eric Vernier}
\affiliation{Laboratoire de Probabilités, Statistique et Modélisation \& CNRS, Université Paris Cité, Sorbonne Université Paris, France}

\maketitle
	

\section{Introduction}
\label{sec:intro}

Quantum computation aims at solving problems which are intractable by classical computers~\cite{nielsen2010quantum}. Yet, simulable quantum circuits (namely, circuits that can be simulated efficiently by classical computers) play an important role in quantum information theory, representing simple models to analyze aspects of information processing, and providing rare benchmarks for testing experimental implementations. In fact, simulable quantum circuits have been known and studied for a long time, with the Clifford ~\cite{gottesman1997stabilizer,gottesman1998theory,gottesman1998heisenberg,aaronson2004improved} and matchgate circuits~\cite{terhal2002classical,vanDenNest2011simulating,brod2016efficient} standing as prominent examples. Now, the emergence of quantum processors of increasingly large scales~\cite{arute2019quantum, bluvstein2022quantum, daley2022practical, kim2023evidence} motivates the search and study of additional simulable models. 

In the past few years, unexpected progress in this direction came from work focusing on quantum circuits as new tractable models for many-body systems out of equilibrium~\cite{fisher2023random,bertini2025exactly}. These studies identified new types of circuits whose dynamics can be either computed exactly or simulated efficiently, with examples including the so-called dual-unitary~\cite{bertini2019exact,rather2020creating,piroli2020exact,claeys2021ergodic,suzuki2022computational,yu2023hierarchical}, Yang-Baxter-integrable~\cite{vanicat2018integrable,ljubotina2019ballistic,aleiner2021bethe,giudice2022temporal,claeys2022correlations,miao2023integrable,vernier2023integrable,vernier2024strong,hubner2025generalized,paletta2025integrability,paletta2025integrability_2}, and automata circuits~\cite{iaconis2019anomalous,iaconis2021quantum,klobas2021exact,pizzi2022bridging,bertini2024exact_east,bertini2025quantum,mazzoni2025breaking}. The results obtained in this literature have contributed to advancing our understanding of several nonequilibrium phenomena, establishing solvable and simulable quantum circuits as a fruitful common playground between quantum information and many-body physics~\cite{fisher2023random,bertini2025exactly}. 

In this context, an interesting recent development has been the discovery of new solvable quantum circuits featuring \emph{free fermions in disguise}~\cite{fendley2019free,fukai2025quantum} (FFD). The notion of FFD was first introduced by Paul Fendley~\cite{fendley2019free}, who presented a spin-chain Hamiltonian with a free fermionic spectrum that cannot be diagonalized by a Jordan-Wigner (JW) transformation~\cite{fendley2019free} (see Refs.~\cite{fendley2007cooper,de2016integrable,feher2019curious} for earlier work in this direction). This fact was later formalized in Ref.~\cite{elman2021free}, highlighting fundamental differences from other models which can be solved via generalizations of the JW transformation~\cite{fradkin1989jordan, wigner1991, huerta1993bose, batista2001generalized, verstraete2005mapping, kitaev2006anyons, nussinov2012arbitrary, chen2018exact, backens2019jordan, tantivasadakarn2020jordan, chapman2020characterization, minami2016solvable, minami2017infinite, yanagihara2020exact, ogura2020geometric, wang2025particle}. The term FFD is now used to identify Fendley's model and later generalizations~\cite{alcaraz2020free,alcaraz2020integrable,fendley2024free,fukai2025quantum,fukai2025free,vernier2025hilbert}.

The problem of finding quantum circuits featuring FFD is very natural but generally hard, due to the fact that Fendley's mapping between spin and fermionic bilinear operators is non-linear and non-local. This is in stark contrast with the JW mapping, which allows one to easily find two-qubit gates (called matchgates~\cite{terhal2002classical,vanDenNest2011simulating,brod2016efficient}) that are individually mapped to fermionic Gaussian operators~\cite{bravyi2004lagrangian}, thus defining quantum circuits with free-fermionic spectrum~\footnote{Note that there also exist quantum circuits that can be mapped to free fermions via the JW mapping, although they are not built out of matchgates, cf. Ref.~\cite{hillberry2024integrability}.}.

In this work, we will build upon the recent results of Ref.~\cite{fukai2025quantum} and demonstrate the FFD solvability of different families of local quantum circuits. Ref.~\cite{fukai2025quantum} provided a first crucial step, by formulating a series of conjectures about the FFD solvability of certain families of quantum circuits. As our first main result, we develop a systematic approach to prove these conjectures, establishing the FFD solvability of quantum circuits with both staircase and brickwork architectures.

Going further, we will explicitly address the question of simulability of FFD circuits. This question is particularly important, as the stastistical-mechanics notion of solvability (typically referring to the possibility of analytically diagonalizing the Hamiltonian) is distinct and does not necessarily imply simulability. As our second main result, we show that the discrete dynamics of certain local observables can be simulated efficiently by classical computers, providing a first step towards the understanding of the quantum and classical complexity of FFD models.

Before leaving this section, we mention that, while finalizing this work, Ref.~\cite{fukai2025free} appeared on the arXiv. There, the authors prove that a special staircase circuit introduced in Ref.~\cite{fukai2025quantum} is solvable with a FFD spectrum. Compared to Ref.~\cite{fukai2025free}, our approach is more general, allowing us to prove \emph{all} the conjectures of Ref.~\cite{fukai2025quantum} (including the FFD solvability of the brickwork circuits).

The rest of this work is organized as follows. In Sec.~\ref{sec:circuits_def} we introduce the circuits whose FFD solvability was conjectured in Ref.~\cite{fukai2025quantum}. Sec.~\ref{sec:transfermatrices} presents the main building blocks of our construction, defining the circuit transfer matrices and proving their commutation relations. These building blocks are used to establish the free fermionic spectrum in Sec.~\ref{sec:solution}. Next, Sec.~\ref{sec:quench_dynamics} is devoted to the study of the circuit dynamics, while our conclusions are consigned to Sec.~\ref{sec:outlook}. Finally, some technical details are provided in Appendix \ref{sec:appendixfermionic}.

\section{The quantum circuits}
\label{sec:circuits_def}

In the original work \cite{fendley2019free}, Fendley considers a local Hamiltonian
\begin{equation}\label{eq:Fendley_ham}
    H = \sum_{m = 1}^M b_m h_m\,,
\end{equation}
defined on a one-dimensional chain of $M$ spins (or qubits). Here, $b_m\in \mathbb{R}$ are arbitrary real numbers, while each $h_m$ is an operator supported on a neighborhood of qubit $m$. The operators $h_m$ satisfy the so-called FFD algebra
\be
\begin{split}
(h_m)^2 &= 1 \,, \\
\{h_m,h_{m+1}\} &= \{h_m,h_{m+2}\}=0 \,,\\
[h_m,h_{l}]&=0\,, \qquad |m-l|>2\,.
\end{split}
\label{eq:FFDalgebra}
\ee
Ref. ~\cite{fendley2019free} demonstrated that the Hamiltonian~\eqref{eq:Fendley_ham} can be completely diagonalised in terms of free fermionic modes, purely based on the algebraic relations above and irrespective of the choice of $b_m$. Note that, contrary to the conventions of Ref. \cite{fendley2019free}, we do not absorb the inhomogeneous couplings in the $h_m$.

Different representations of the algebra~\eqref{eq:FFDalgebra} give rise to different models. While the exact solution presented in the next sections only relies on the algebraic properties and does not depend on the representation, we will focus on a specific one. Following~\cite{fendley2019free}, we set
\be
h_m = Z_{m-2} Z_{m-1} X_m  \,,\qquad m=1\ldots M \,,
\label{representation} 
\ee 
where $X_m$ and $Z_m$ are Pauli matrices acting on spin $m$ and as the identity elsewhere, with the convention that $X_{m}=Y_m=\openone$ for $ m\leq 0$.

The family of circuits that we consider are defined in terms of the local unitary gates
\be 
g_m = \cos\frac{\phi_m}{2} +  i \sin \frac{\phi_m}{2} h_m \,, 
\label{eq:gi}
\ee
where the $\phi_m$ are real parameters, controlling the spatial inhomogeneity of the circuit, and playing a role similar to the couplings $b_m$ for the Hamiltonian model. Due to the algebraic construction presented below, their values can be chosen arbitrarily, and we only fix them when considering the numerical implementation of the dynamics. We now present three circuit architectures patching these gates in different space-time patterns.  These circuits were introduced in Ref.~\cite{fukai2025quantum}, which first conjectured their FFD solvability. Throughout this paper, we will focus on representations for which $g_m=g_m^{\rm T}$, as in \eqref{representation}, but our discussion can easily be adapted to the more general settings considered in \cite{fukai2025quantum}, where $g_i$ and $g_i^{\rm T}$ are not necessarily equal. 

\begin{figure}
    \centering
    \includegraphics[scale=1.7]{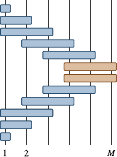}\qquad
    \includegraphics[scale=1.7]{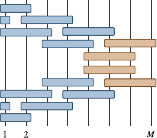}\qquad
    \includegraphics[scale=1.7]{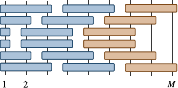}
    \caption{Pictorial representation of the circuits \eqref{eq:circuitI}, \eqref{eq:circuitII} and \eqref{eq:circuitIII}: the circuits are built up from the local gates $g_m$ [Eq. \eqref{eq:gi}], spanning the sites $\{m-2,m-1,m\}$  (the color distinction only serves as a guide for the eye for Section \ref{sec:transfermatrices}). By definition, $g_1$ and $g_2$ only act on 1 and 2 sites, respectively. Notice how contrary to circuits \eqref{eq:circuitI} and \eqref{eq:circuitII}, the depth of circuit \eqref{eq:circuitIII} doesn't scale with the system size $M$.}
    \label{fig:circuits}
\end{figure}

The first family of circuits corresponds to the evolution operator 
\be
\mathcal{V}^{({\rm I})}_M = G_M \cdot G_M^T \,,  \qquad G_M =g_1 g_2 \ldots g_M \,,
\label{eq:circuitI}
\ee 
defining the discrete dynamics $\ket{\psi_t}=\left(\mathcal{V}^{({\rm I})}_M\right)^t\ket{\psi_0}$ for $t\in \mathbb{N}$. 
The second family is defined for even values of $M$, and is associated to a discrete evolution operator with two-site periodicity:
\begin{align}
\mathcal{V}^{({\rm II})}_M =& 
 G_M \cdot G_M^T \,, \qquad G_M= 
(g_2 g_4  \ldots g_{M})(g_1 g_3  \ldots g_{M-1})
\,.
\label{eq:circuitII}
\end{align}
The FFD solvability of these two families of quantum circuits has been already established in the literature. First, although Ref.~\cite{fendley2019free} only studied Hamiltonian models, $\mathcal{V}^{({\rm I})}_M$ coincides with a transfer-matrix operator introduced in that work and its FFD solvability easily follows from the constructions therein. The circuits defined by $\mathcal{V}^{({\rm II})}_M$ were instead introduced in Ref.~\cite{fukai2025quantum} and later shown to be FFD solvable in Ref.~\cite{fukai2025free}.

The third and last family we shall consider has a ``brickwork'' evolution operator with three-site periodicity: 
\begin{align}
\mathcal{V}^{({\rm III})}_M &= G_M \cdot G_M^T \,,
\nonumber\\ 
G_M &=(g_3 g_6  \ldots g_{M})
(g_2 g_5  \ldots g_{M-1})
 (g_1 g_4  \ldots g_{M-2})
\,,
\label{eq:circuitIII}
\end{align}
(accordingly, it is defined for $M$ multiple of $3$). Contrary to the first two cases, $\mathcal{V}^{({\rm III})}_M$ can be written as a ``finite-depth'' circuit. That is, its application only requires a time which is independent of the system size, since gates acting on disjoint sets of qubits can be applied in parallel. Such circuits were also introduced in Ref.~\cite{fukai2025quantum} and conjectured to be FFD solvable, but no proof or exact solution has been presented so far. 

The three families of circuits are represented in Figure \ref{fig:circuits}, where the rectangles denote the gates $g_m$, with the gate for $g_m$ spanning sites from $m-2$ to $m$: in the representation \eqref{representation} this is in accordance with the definition of the gates, which act non-trivially on three consecutive spins only.

In the next sections, we will present an exact solution of the three families of circuits \eqref{eq:circuitI}, \eqref{eq:circuitII} and \eqref{eq:circuitIII}.  
As mentioned above, the solution relies only on the FFD algebra \eqref{eq:FFDalgebra}, and not on the specific representation. More precisely, we will make extensive use of the relations
\be 
\begin{split}
(g_m)^2 &=  x_m + i y_m h_m  \, ,\\
g_m h_{m\pm a} g_m &=  h_{m\pm a} \,,\qquad a=1,2 \,,
\label{galgebra}
\end{split}
\ee 
which can be easily derived from~\eqref{eq:FFDalgebra}. Here, we have introduced the shorthand notations $x_m=\cos\phi_m$ and $y_m= \sin\phi _m$.

\section{Commuting transfer matrices}
\label{sec:transfermatrices}

We now present our solutions of the circuits. The first step in each case is to embed the evolution operator in a family of mutually commuting transfer matrices. 
While this can be achieved through several routes (see Ref. \cite{fendley2019free} for circuit I, and \cite{fukai2025free} for circuit II), the approach presented here treats all three circuits on the same footing. It starts by evaluating the evolution operators ``from right to left'', as we now describe starting with circuit I. 

\subsection{Warmup : circuit I}

Using the algebra \eqref{galgebra}, we can evaluate the rightmost pair of gates, $(g_M)^2$ (represented in orange on Fig. \ref{fig:circuits}) as a sum of two terms, one proportional to the identity and one proportional to $h_M$. For the first term, we can then evaluate the next product $(g_{M-1})^2$ similarly, while for the second term $g_{M-1} h_M g_{M-1}$ can be evaluated using the second line of \eqref{galgebra}. We can iterate this procedure, noticing that at each step the evaluation of the product of two $g_m$ only depends on whether $h_{m+1}$ and/or $h_{m+2}$ are present. We then find convenient to introduce the following four families of operators
\begin{align} 
A_m &= G_m G_m^T  \,, \label{eq:Adef} \\
B_m &= i~G_m h_{m+1} G_m^T \,, \label{eq:Bdef} \\ 
C_m &= i~G_m h_{m+2} G_m^T \,,  \label{eq:Cdef} \\
D_m &= G_m h_{m+1}h_{m+2} G_m^T  \label{eq:Ddef} \,, 
\end{align} 
where, similar to Eq.~\eqref{eq:circuitI}, $G_m=g_1 g_2 \ldots g_m$. By computing the product of the two $g_m$ operators in Eqs.~\eqref{eq:Adef},~\eqref{eq:Bdef}, and~\eqref{eq:Cdef} according to the algebra \eqref{galgebra}, it is easy to show that these operators satisfy the recursion relations
\begin{align}
A_{m} &= x_m A_{m-1} + y_m B_{m-1} \,,\label{recursionAI}  \\
B_{m} &=  C_{m-1} \label{recursionBI}\,, \\
C_{m} &=  i~A_{m-1} h_{m+2}  \label{recursionCI} \,.
\end{align}
Note that $D_m$ does not enter the recursion relations and could have been ignored at this stage. However we introduce it as it will be needed for the other families of circuits.

The recursion relations~\eqref{recursionAI},~\eqref{recursionBI}, and~\eqref{recursionCI} allow us to express the evolution operator $\mathcal{V}_M^{({\rm I})}=A_M$ as a matrix product operator (MPO)~\cite{cirac2021matrix}. This idea is not new: in Ref.~\cite{fendleycp}, Fendley has used this strategy to rewrite the transfer matrices of free-parafermionic models as two-dimensional MPOs, exploiting an algebra very similar to \eqref{galgebra}. In the present case the MPO has a three-dimensional ancillary space. We denote the basis states by $|a\rangle,|b\rangle,|c\rangle$, in $1$-to-$1$ correspondence with the operators $A,B,C$ above. The operator $\mathcal{V}_M^{({\rm I})}=A_M$ then takes the form
\be 
\mathcal{V}_M^{\rm (I)} = (1~1~1)~\Omega_1 \Omega_2 \ldots \Omega_M ~|a\rangle \,,
\label{eq:ABCIMPO}
\ee 
where the matrices $\Omega_m$ encode the recursion relations as
\be 
\Omega_m = \left(  \begin{array}{ccc}  x_m & 0 & 1 \\ i y_m h_m & 0 & 0 \\ 0 & 1 & 0  \end{array} \right) \,.
\ee 

It is known from Fendley's original study of FFD \cite{fendley2019free} that operators with the geometry \eqref{eq:circuitI} can be embedded in a family of mutually commuting transfer matrices, which can be constructed as generating functions of mutually commuting charges of increasing order in the $h_m$. To recover these charges, we therefore promote the MPOs \eqref{eq:ABCIMPO} to families $A_m(u)$, $B_m(u)$, $C_m(u)$ depending on a \emph{spectral parameter} $u$, by simply attaching a factor $u$ to each $h_m$ appearing in the MPOs. 
The resulting one-parameter families read 
\begin{align} 
 A_m(u)&= (1~1~1)~\Omega_1(u) \Omega_2(u) \ldots \Omega_m(u) ~|a\rangle  \,,
 \label{eq:AuIMPO}
 \\
 B_m(u)&= (1~1~1)~\Omega_1(u) \Omega_2(u) \ldots \Omega_m(u) ~|b\rangle  i u h_{m+1}  \,,
 \label{eq:BuIMPO}
 \\
C_m(u)&= (1~1~1)~\Omega_1(u) \Omega_2(u) \ldots \Omega_m(u) ~|c\rangle  i u h_{m+2} \,,
\label{eq:CuIMPO}
\end{align} 
where the local matrices $\Omega_m(u)$ now depend upon the spectral parameter $u$ through
\be 
\Omega_m(u) = \left(  \begin{array}{ccc}  x_m & 0 & 1 \\ i u y_m h_m & 0 & 0 \\ 0 & 1 & 0  \end{array} \right) \,.
\label{eq:OmegaI}
\ee 
It can indeed be checked that the $A_m(u)$ defined in this way (which, we stress, recover the circuit evolution operator \eqref{eq:ABCIMPO} as $\mathcal{V}_M^{\rm (I)}=A_M(1)$), commute with one another. This fact is easily seen by including their commutation relations into a more general set of relations, involving the families $A,B,C$, reading 
\begin{align}
&[A_m(u),A_m(v)]=[B_m(u),B_m(v)]=[C_m(u),C_m(v)]=0 \,,
\nonumber \\ 
&[A_m(u),B_m(v)] + [B_m(u),A_m(v)]  = 0\,,
\nonumber\\
&[A_m(u),C_m(v)] + [C_m(u),A_m(v)]  = 0\,,
\nonumber\\
&[B_m(u),C_m(v)] + [C_m(u),B_m(v)]  = 0\,,
\nonumber\\
&u \{A_m(u),B_m(v)\} =  v \{B_m(u),A_m(v)\}  \,,
\nonumber\\
& u \{A_m(u),C_m(v)\} =  v \{C_m(u),A_m(v)\}  \,.
\label{eq:ABCalgebra}
\end{align}
The algebra \eqref{eq:ABCalgebra} can be proved recursively from the relations 
\begin{align}
A_{m}(u) &= x_m A_{m-1}(u) + y_m B_{m-1}(u) \label{recursionAIu} \,, \\
B_{m}(u) &=   C_{m-1}(u) \,,\label{recursionBIu} \\
C_{m}(u) &= i u x_m  A_{m-1}(u) h_{m+2}\,,  \label{recursionCIu} 
\end{align}
[which are again easily derived from the definitions \eqref{eq:AuIMPO}, \eqref{eq:BuIMPO}, \eqref{eq:CuIMPO}] together with the initial values $A_0(u)=1$, $B_0(u)=i u h_1$, $C_0(u)=i u h_2$.

We will now see that the same algebra (extended by the inclusion of a fourth family of operators) appears when studying the circuits II and III. In Sec.~\ref{sec:solution}, it will be used to derive the FFD spectrum of the circuits.

\subsection{Circuits II and III}

For circuits II and III we follow the same strategy as described above, namely we start by evaluating the evolution operator from right to left, and recast the result in a MPO form. 
Defining operators $A_m$, $B_m$, $C_m$, $D_m$ exactly as before, namely Eqs.~\eqref{eq:Adef},~\eqref{eq:Bdef},~\eqref{eq:Cdef}, and~\eqref{eq:Ddef}, we find recursion relations which now involve all four families instead of just three, resulting in MPOs with a four-dimensional ancillary space. 
Because of the spatial structure of the circuits, the recursion relations relate operators $A,B,C,D$ at a given $m$ to those at $m-2$ (resp. $m-3$), corresponding to evaluating the products of local gates represented in orange in Fig.~\ref{fig:circuits}.
The resulting MPO forms of the evolution operators then read 
\begin{align} 
\mathcal{V}_M^{\rm (II)}&=
(1~1~1~1)~\Omega_1^{\rm (II)} \Omega_3^{\rm (II)} \ldots \Omega_{M-1}^{\rm (II)} ~|a\rangle \,,
\label{eq:ABCIIMPO}
\\
\mathcal{V}_M^{\rm (III)}&=
(1~1~1~1)~\Omega_1^{\rm (III)} \Omega_4^{\rm (III)} \ldots \Omega_{M-2}^{\rm (III)} ~|a\rangle \,,
\label{eq:ABCIIIMPO}
\end{align} 
where we have introduced the canonical basis  $|a\rangle$, $|b\rangle$, $|c\rangle$, $|d\rangle$ of the ancillary space, while the local matrices $\Omega_m$ read 
\begin{widetext}
\begin{align}
\Omega_m^{\rm (II)}&=
\Theta_m \cdot   \left(\begin{array}{cccc}
x_m x_{m+1} & 1 & x_{m} & x_{m+1} \\
 1 & 0 &   x_{m+1} & 0 \\ 
 x_m  & 0 & 0 &    1 \\
0 & 0 & 1 & 0
 \end{array}   \right)\,,
 \\
\Omega_m^{\rm (III)}&=
\Theta_m \cdot  \left(\begin{array}{cccc}
\kappa^+_{m+2} x_m x_{m+1} & x_{m} & x_m x_{m+1} & x_m\kappa^+_{m+2} \\
1   &\kappa^-_{m+2}x_{m+1} &  \kappa^-_{m+2}  & x_{m+1} \\ 
x_{m} & 0 &\kappa^-_{m+2}x_m  & 0 \\
0 & 1 & 0 & \kappa^+_{m+2}  
 \end{array}   \right)
 \,.
\end{align} 
\end{widetext}
Here, we have introduced the shorthand notation $\kappa_{m+2}^\pm = x_{m+2} \pm i y_{m+2} h_{m+2} $, and we have defined $\Theta_m = \mathrm{diag}(1,i y_m h_m, i y_{m+1} h_{m+1} , y_{m} y_{m+1} h_{m} h_{m+1})$.

Following the case of circuit I, we want to lift the MPOs~\eqref{eq:ABCIIMPO} and~\eqref{eq:ABCIIIMPO} to families of mutually commuting transfer matrices. 
A naive guess would be to put a spectral parameter $u$ in front of each $h_m$ appearing in the MPOs. This procedure, however, does not yield commuting operators, as can be checked numerically for small system sizes. 
Instead, we find, by inspection, that commuting transfer matrices are obtained by applying the following rule to the MPOs: inside each $\Omega_m$, linear and trilinear terms in the densities $\{h_m\}$ are multiplied by a $u$, while bilinear terms are not. In formulas, this can be expressed by promoting the $\Omega_m$ to the following $u$-dependent matrices 
\begin{widetext}
\begin{align}
\Omega_m^{\rm (II)}(u)&=
\Theta_m \cdot   \left(\begin{array}{cccc}
x_m x_{m+1} & 1 & x_{m} & x_{m+1} \\
 u & 0 &   u x_{m+1} & 0 \\ 
 u x_m  & 0 & 0 &    u \\
0 & 0 & 1 & 0
 \end{array}   \right)\,,
 \\
\Omega_m^{\rm (III)}(u)&=
\Theta_m \cdot  \left(\begin{array}{cccc}
\kappa^+_{m+2}(u) x_m x_{m+1} & x_{m} & x_m x_{m+1} & x_m\kappa^+_{m+2}(u) \\
u   &\kappa^-_{m+2}(u)x_{m+1} &  \kappa^-_{m+2}(u)  & u  x_{m+1} \\ 
u x_{m} & 0 &\kappa^-_{m+2}(u)x_m  & 0 \\
0 & 1 & 0 & \kappa^+_{m+2}(u)  
 \end{array}   \right)
 \,, 
\end{align} 
\end{widetext}
where on the second line we introduced  
\begin{align} 
\kappa_{m+2}^+(u)&= x_{m+2} + i u y_{m+2} h_{m+2}\,, \\ 
{\kappa}^-_{m+2}(u)&= u x_{m+2} - i y_{m+2} h_{m+2}
\,.
\label{kappa}
\end{align} 

As in the previous section, commutation of transfer matrices is most easily shown by introducing (now four) families of MPOs satisfying the algebra \eqref{eq:ABCalgebra} (plus additional relations). The first three families, $A_m(u)$, $B_m(u)$, $C_m(u)$ are defined analogously as for circuit I, see Eqs.~\eqref{eq:AuIMPO}, \eqref{eq:BuIMPO}, and ~ \eqref{eq:CuIMPO}, with the difference that $m$ must be a multiple of $2$ (resp. $3$) for circuit II (resp. III), and that only the matrices $\Omega_m$ corresponding to every other (resp. third) site appear in the MPO.
We similarly define the families $D_m(u)$ as   
\begin{align} 
D_m(u)=\begin{cases}
(1~1~1~1)~\Omega^{\rm (II)}_1(u) \Omega^{\rm (II)}_3(u) \ldots \Omega^{\rm (II)}_m(u) ~|d\rangle  h_{m+1} h_{m+2} \qquad \mbox{for circuit II,}
\\
(1~1~1~1)~\Omega^{\rm (III)}_1(u) \Omega^{\rm (III)}_4(u) \ldots \Omega^{\rm (III)}_m(u) ~|d\rangle  h_{m+1} h_{m+2}  \qquad \mbox{for circuit III.}
\end{cases}
\label{eq:DuIIandIIIMPO}
\end{align} 

 Again, it follows from the definition that $\mathcal{V}_M^{\rm (II,III)} = A_M(1)$.
For both circuits, we find that the $A$,$B$,$C$, and $D$ operators satisfy the same algebra derived for circuit I, cf. Eq.~\eqref{eq:ABCalgebra}, together with the additional set of relations 
\begin{align} 
& [D_m(u),D_m(v)]=0 \,,
\nonumber \\
& 
 \{A_m(u),D_m(v)\} =   \{D_m(u),A_m(v)\} \,.
 \label{eq:Dalgebra}
\end{align}
Once again, the algebraic relations \eqref{eq:ABCalgebra} and \eqref{eq:Dalgebra} can be proved iteratively, using a set of recursion relations obeyed by the $A_m(u)$, $B_m(u)$, $C_m(u)$, $D_m(u)$ which follows directly from their MPO expression. 
For circuit II these are
\begin{align}
    A_{m}(u) &= x_m x_{m-1}A_{m-2}(u) + y_{m-1} B_{m-2}(u) + x_{m-1} y_m C_{m-2}(u)\,, \label{recursionAII}  \\
B_{m}(u) &=   i u A_{m-2}(u) h_{m+1}\,, \label{recursionBII} \\
C_{m}(u) &= i u (x_{m-1}  A_{m-2}(u) + x_m y_{m-1} B_{m-2}(u) + y_{m-1} y_m D_{m-2}(u) )h_{m+2}\,,  \label{recursionCII} \\ 
D_m(u) &= (x_m A_{m-2}(u) + y_m C_{m-2}(u)) h_{m+1} h_{m+2}\,,
\label{recursionDII}
\end{align}
while for circuit III, they are 
\begin{align}
    A_{m}(u) &= x_{m-1} x_{m-2}A_{m-3}(u)\kappa^+_m(u) + y_{m-2} B_{m-3}(u) + x_{m-2} y_{m-1} C_{m-3}(u)\,, \label{recursionAIII}  \\
B_{m}(u) &=   i [u x_{m-2} A_{m-3}(u) + x_{m-1}y_{m-2}B_{m-3}(u)\kappa^-_m(u) + u y_{m-2} y_{m-1} D_{m-3}(u)] h_{m+1}\,, \label{recursionBIII} \\
C_{m}(u) &= i [u x_{m-1} x_{m-2}  A_{m-3}(u) \!+\! (y_{m-2} B_{m-3}(u) + x_{m-2} y_{m-1}  C_{m-3}(u)) \kappa^-_m(u) ]h_{m+2}\,,  \label{recursionCIII} \\ 
 D_m(u) &= [(x_{m-2} A_{m-3}(u)\!+\! y_{m-1}y_{m-2} D_{m-3}(u))\kappa^+_m(u) \!+\! y_{m-2} x_{m-1} B_{m-3}(u)] h_{m+1} h_{m+2}.
\label{recursionDIII}
\end{align}

\subsection{Conserved charges}
\label{sec:charges}

As a result of the constructions presented so far, we have established that, for all circuits I, II and III, the time evolution generator $\mathcal{V}_M = A_M(1)$ commutes with a continuous family of transfer matrices, 
\be 
[\mathcal{V}_M , A_M(u) ] = 0 \,.
\ee 
As usual in the study of integrable models~\cite{vanicat2018integrable}, such transfer matrices can be used to generate conserved charges that are left invariant by the dynamics. In our case, this is done by expanding $A_M(u)$ in powers of $u$, around the point $u=0$, as we now explain.

First, in all cases, it is easy to check that at $u=0$ the transfer matrices are proportional to the identity, namely
\be 
A_M(0) = \prod_{m=1}^M x_m  \,.
\ee 
Noticing that $A(i u)$ is an hermitian operator for all $u\in \mathbb{R}$, we define the first hermitian conserved charge ("the Hamiltonian") as
\be 
H= \left. \frac{\mathrm{d}}{\mathrm{d}u} \ln A_M(i u) \right|_{u=0} \,.
\ee 

For circuit I, it is of the form
\be 
H^{\rm (I)} = \sum_{m=1}^M b_m h_m \,, \qquad b_m \equiv \frac{y_m}{x_{m-2}x_{m-1}x_m} \,,
\label{eq:HI}
\ee 
where by convention $x_m=1$ for $m\leq 0$. This is the Hamiltonian originally studied by Fendley \cite{fendley2019free}. The higher order charges, which are of increasing order in the $h_m$, can similarly be recovered by taking higher logarithmic derivatives of $A_M(iu)$.

For circuit II, we find that the first charge is of the form  
\be 
H^{\rm (II)} = \sum_{m=1}^M b_m h_m + \sum_{m=1}^M b_{m,m+1,m+3} h_m h_{m+1} h_{m+3} \,,
\ee 
where now 
\begin{align} 
b_m &= \begin{cases} 
\frac{y_m}{x_{m-2} x_m} \qquad \mbox{if $m$ even,} \\
\frac{y_m}{x_{m-2} x_{m-1 }x_m x_{m+1}} \qquad \mbox{if $m$ odd,}
\end{cases}
\\
b_{m,m+1,m+3} &= \frac{y_m y_{m+1} y_{m+3}}{x_{m-2} x_m x_{m+1} x_{m+3}}  \,.
\end{align} 
In the homogeneous case where all $\phi_m$ are equal, this Hamiltonian coincides (up to a mirror transformation) with the one found in Ref.~\cite{fukai2025quantum} via a brute-force approach (namely, by fixing the terms in the Hamiltonian imposing commutation with the circuit evolution operator for finite size).
Similarly, considering higher transfer matrices yields charges of higher order in the $h_m$, which we will not describe here.

The case of circuit III is different, as here the first logarithmic derivative of $A_M(u)$ involves terms of order increasing with the size $M$ (as it can be checked numerically), and therefore does not exhibit the locality properties of the Hamiltonians found for circuits I and II. It remains an open question whether local charges for the circuit III can be recovered from our algebraic constructions.

\section{The free fermionic spectrum}
\label{sec:solution}

In the previous section, we have shown that, in all three geometries, the time evolution generator $\mathcal{V}_M$ can be identified as an element of a mutually commuting family of operators $A_M(u)$. In turn, we have shown that $A_M(u)$ is part of a larger algebra involving additional families of operators, $B_M(u)$, $C_M(u)$, and $D_M(u)$ (the latter being necessary only for circuits II and III).

As we will now see, these operators and their algebra are enough to compute the spectrum of the evolution operator, which will turn out be of free-fermionic form, and to construct the corresponding raising and lowering operators. 
The construction follows very closely Fendley's original solution \cite{fendley2019free}, with the advantage that it treats all circuits on a common footing. 

Our construction starts with the observation that, in all three cases, the products $\mathcal{A}_m(u) = A_m(u) A_m(-u)$ (and similarly for $B$,$C$, and $D$) are proportional to the identity. 
This fact can be shown very simply through a set of recursion relations for the operators $\mathcal{A},\mathcal{B},\mathcal{C}$, and $\mathcal{D}$. These recursion relations are derived by combining the ones of $A$, $B$, $C$, and $D$, together with the algebraic relations \eqref{eq:ABCalgebra}, and \eqref{eq:Dalgebra}. 
For circuit I, the recursion relations read 
\begin{align}
\mathcal{A}_{m}(u)&=x_m^2 \mathcal{A}_{m-1}(u) + y_m^2 \mathcal{B}_{m-1}(u)\,,\\
\mathcal{B}_{m}(u)&= \mathcal{C}_{m-1}(u)\,,  \\
\mathcal{C}_{m}(u)&= u^2  \mathcal{A}_{m-1}(u)  \,.
\end{align}
Note that these relations immediately imply $\mathcal{A}_m(u) = x_m^2\mathcal{A}_{m-1}(u) + u^2 y_m^2 \mathcal{A}_{m-3}(u)$, recovering the recursion relation derived in Ref.~\cite{fendley2019free}. For circuit II, they take the form
\begin{align}
\mathcal{A}_{m}(u)&=x_{m-1}^2 x_m^2 \mathcal{A}_{m-2}(u) + y_{m-1}^2 \mathcal{B}_{m-2}(u) + x_{m-1}^2 y_m^2 \mathcal{C}_{m-2}(u)\,,\\
\mathcal{B}_{m}(u)&= u^2 \mathcal{A}_{m-2}(u)\,,  \\
\mathcal{C}_{m}(u)&= u^2 \left( x_{m-1}^2 \mathcal{A}_{m-2}(u) + x_m^2 y_{m-1}^2 \mathcal{B}_{m-2}(u) - y_m^2 y_{m-1}^2 \mathcal{D}_{m-2}(u) \right)\,, 
\\
\mathcal{D}_m(u) &= -x_m^2 \mathcal{A}_{m-2}(u) - y_m^2 \mathcal{C}_{m-2}(u) \,,
\end{align}
and for circuit III,
\begin{align}
\mathcal{A}_{m}(u)&=x_{m-2}^2 x_{m-1}^2 (x_m^2+u^2 y_m^2) \mathcal{A}_{m-3}(u) + y_{m-2}^2 \mathcal{B}_{m-3}(u) + x_{m-2}^2 y_{m-1}^2 \mathcal{C}_{m-3}(u)\,, \\
\mathcal{B}_{m}(u)&= u^2 x_{m-2}^2 \mathcal{A}_{m-3}(u) +y_{m-2}^2 x_{m-1}^2  (u^2 x_m^2 + y_m^2) \mathcal{B}_{m-3}(u) - u^2 y_{m-1}^2 y_{m-2}^2 \mathcal{D}_{m-3}(u)\,, 
\\
\mathcal{C}_{m}(u)&= u^2  x_{m-2}^2 x_{m-1}^2 \mathcal{A}_{m-3}(u) +\left( y_{m-2}^2  \mathcal{B}_{m-3}(u) + x_{m-2}^2 y_{m-1}^2 \mathcal{C}_{m-3}(u) \right)(u^2 x_m^2 + y_m^2)\,, 
\\
\mathcal{D}_m(u) &= (-x_{m-2}^2 \mathcal{A}_{m-3}(u) + y_{m-2}^2 y_{m-1}^2 \mathcal{D}_{m-3}(u))(x_m^2 + u^2 y_m^2) - y_{m-2}^2 x_{m-1}^2 \mathcal{B}_{m-3}(u)  \,.
\end{align}

Now, when supplemented with the initial conditions
\be 
\mathcal{A}_0(u) = 1 \,, \qquad \mathcal{B}_0(u)=\mathcal{C}_0(u)= u^2 \,, \qquad 
\mathcal{D}_0(u)= -1 \,,
\ee 
these recursion relations imply that $\mathcal{A}_m(u)$, $\mathcal{B}_m(u)$, $\mathcal{C}_m(u)$, and $\mathcal{D}_m(u)$ are proportional to the identity for all $m$. More precisely, in all three geometries $\mathcal{A}_M(u)$ is a polynomial in $u^2$ of degree 
\be 
S= \left\lfloor \frac{M+2}{3}  \right\rfloor \,,
\ee 
 with $S$ pairs of opposite purely imaginary roots. Labeling the latter as 
\be 
\pm i u_1 \,, \pm i u_2 \,, \ldots \pm i u_S \,, 
\ee 
we can rewrite 
\be 
\mathcal{A}_M(u) = A_M(u) A_M(-u) =\prod_{k=1}^S \frac{u^2 + u_k^2}{1+ u_k^2} \,, 
\ee 
where the normalization has been fixed by noticing that $\mathcal{A}_M(1)=1$ in all three geometries.

The roots of the polynomial $\mathcal{A}_M$ encode all the eigenvalues of the transfer matrices $A_M(u)$, and therefore of the evolution operator $\mathcal{V}_M=A_M(1)$. Following the lines of Fendley's original work~\cite{fendley2019free}, this is best seen by constructing explicitly the corresponding fermionic creation/annihilation operators. To this end one needs to define an additional boundary mode $\chi$ squaring to the identity and which commutes with all $h_m$ except at the boundary: for circuits I and III one needs to take 
\be 
\chi^2= 1 \,, \qquad \{\chi,h_M\}=0 \,, \qquad [\chi,h_m]=0\quad\mbox{for $m<M$} \,, 
\label{eq:chiIandIII}
\ee 
while for circuit II one needs instead 
\be 
\chi^2= 1 \,, \qquad \{\chi,h_M\}=\{\chi,h_{M-1}\}=0 \,, \qquad [\chi,h_m]=0\quad\mbox{for $m<M-1$} \,. 
\label{eq:chiII}
\ee 
In the representation \eqref{representation}, a natural choice for such an operator is $\chi=Z_M$ for circuits I and III, and $\chi=Z_{M-1}Z_M$ for circuit II.

Having defined the boundary mode $\chi$, fermionic creation/annihilation operators may then be constructed as 
\be \label{eq:fermion-def}
\Psi_{\pm k} = \frac{1}{N_k} A_M(\pm i u_k) \chi A_M(\mp i u_k)\,,
\ee 
for $k=1\ldots S$, where the normalizations $N_k$, derived in Appendix \ref{sec:appendixfermionic}, are given by 
\be 
(N_k)^2 = \begin{cases} \label{eq:norm}
8 i u_k x_M^2 \mathcal{A}_{M-1}(i u_k) 
\mathcal{A}_M'(-i u_k) \qquad & \mbox{for circuit I,} 
\\ 
8 i u_k ( x_M x_{M-1})^2 \mathcal{A}_{M-2}(i u_k) \mathcal{A}_M'(-i u_k)
\qquad &\mbox{for circuit II,} 
\\
- 8 i u_k   (i u_k x_{M-2} x_{M-1}y_M)^2  \mathcal{A}_{M-3}(i u_k) 
\mathcal{A}_M'(-i u_k) \qquad\qquad & \mbox{for circuit III}\,,
\end{cases}
\ee 
where $\mathcal{A}^\prime_m(u)$ denotes the derivative with respect to the spectral parameter $u$.
In Appendix~\ref{sec:appendixfermionic}, we prove using the algebra of operators $A$, $B$, $C$, and $D$ and their recursion relations that the operators~\eqref{eq:fermion-def} obey canonical anticommutation relations
\be 
\{\Psi_k, \Psi_l\} = \delta_{k+l,0} \,.
\label{eq:canonical}
\ee 
Using similar techniques, we also prove in Appendix~\ref{sec:appendixfermionic} the formula
\be \label{eq:fermion-time-evolution}
(i u_k-u) A_M(u) \Psi_k = (i u_k + u) \Psi_k A_M(u) \,, 
\ee 
which states that the operators $\Psi_k$ act as mode creation/annihilation operators for the transfer matrices $A_M(u)$.
In fact, we find that $A_M(u)$ can be fully expressed in terms of the corresponding occupation numbers as 
\be 
A_M(u) =  \prod_{k=1}^S \frac{i u_k + u  [\Psi_k , \Psi_{-k}]}{\sqrt{1+u_k^2}}\,.
\label{eq:Aintermsoffermions}
\ee 
In principle, one could establish Eq.~\eqref{eq:Aintermsoffermions} by following the proof of Ref.~\cite{fendley2019free}, which is based on a construction involving the higher charges of the model. This strategy, however, appears cumbersome for the circuits II and III, since the charges are more complicated. Therefore, we have instead established Eq.~\eqref{eq:Aintermsoffermions} numerically, by testing it for many values of $u$ and system sizes up to $M=12$. 

Eq.~\eqref{eq:Aintermsoffermions} allows us to rewrite the evolution operator as a Gaussian operator: setting $u=1$ and introducing the pseudoenergies 
\be 
\epsilon_k = \arctan \frac{1}{u_k}\,,
\ee 
we have
\be \label{eq:diagonal_form}
\mathcal{V}_M =
\exp\left(-   i \sum_{k=1}^S \epsilon_k  [\Psi_k,\Psi_{-k}]   \right) \,. 
\ee 
Eq.~\eqref{eq:diagonal_form} establishes the anticipated FFD spectrum of the model.

\section{Classical simulation of the quantum circuits}
\label{sec:quench_dynamics}

In the past sections, we have shown that the quantum circuits I, II, and III all admit a free-fermionic spectrum, with the explicit diagonal form of the Floquet unitary operator given in Eq.~\eqref{eq:diagonal_form}. As mentioned, this result does not immediately imply that the circuit dynamics can be computed exactly, raising the question of the classical simulability of FFD solvable circuits. In this section, we tackle this question, providing some preliminary results.  We note that the time evolution of FFD solvable models was also studied in Ref.~\cite{vona2014exact}. However, that reference focused on  dynamical infinite-temperature correlation functions, not genuine nonequilibrium protocols.

For concreteness, we consider a simple protocol where the system is initialized in a product state, and focus on the dynamics of local observables. This protocol is very natural from a theoretical many-body perspective, mimicking a quantum-quench problem in the Hamiltonian setting~\cite{essler2016quench}. At the same time, the protocol is also natural from the experimental point of view, as current prototypes of quantum simulators are typically initialized in unentangled states and evolved by applying geometrically local gates~\cite{altman2021quantum}. 

Before discussing our results, it is instructive to compare our setting with the case of circuits which are solvable by a JW transformation. In the latter case, it is often trivial to map initial product states into Gaussian fermionic states and, as a consequence, the dynamics can be either solved analytically~\cite{essler2016quench} or simulated efficiently using classical tools from quantum information theory~\cite{bravyi2004lagrangian,surace2022fermionic}. Unfortunately, the dictionary between states in the spin and fermion Hilbert spaces is much more complicated for FFD solvable models, making this approach problematic. Therefore, while recent work has reported progress in this direction~\cite{vernier2025hilbert}, we will follow a different strategy based on simulating the Heisenberg evolution of local observables. 

For concreteness, we focus on the dynamics generated by $\mathcal{V}^{({\rm III})}_M$. This circuit appears to be more suitable for implementation in current noisy quantum devices~\cite{preskill2018quantum}, as the corresponding Floquet unitary operator can be realized as a circuit of finite depth, thus potentially reducing the effect of decoherence. In addition, its brickwork structure guarantees that correlation functions are contained in a lightcone -- models with this property are called quantum cellular automata~\cite{arrighi2019overview,farrelly2020review}, and are particularly interesting as they mimic many features of local-Hamiltonian dynamics.

Our strategy for simulating the dynamics of local observables is very simple. Given a local (spin) observable $\mathcal{O}$, we first expand it in terms of the fermionic modes $\Psi_s$. The latter are trivial to evolve, so that computing the time evolution of $\mathcal{O}$ essentially reduces to the computation of the initial expectation values of the modes. While this is in principle still a complicated task, we show that these expectation values can be computed efficiently, exploiting the MPO form of $\Psi_s$.

Clearly, our approach requires to find the expression of local observables in terms of the fermionic modes. Unfortunately, solving this ``inverse problem'' is not an easy task and it is still unsolved for generic local operators. Luckily, partial results were obtained in Ref.~\cite{vona2014exact}, which identified families of local operators admitting a simple expression in terms of the fermionic modes. In the following, we will benchmark our approach for the simplest of the observables identified in Ref.~\cite{vona2014exact}, namely the edge operator $\chi$.

Following Ref.~\cite{vona2014exact}, we start by assuming the decomposition
\begin{equation}
    \chi = \sum_{s = -S}^{S}c_s\Psi_s\,.\label{eq:chi-expansion}
\end{equation}
Apart from the fermionic modes, the expansion includes a Majorana mode $\Psi_0$ defined through \cite{vona2014exact}
\begin{equation}\label{eq:psi_0_def}
\mathcal{Q} = c_0\Psi_0\,,\quad \mathcal{Q} = \lim_{u\rightarrow \infty}\frac{1}{2}\left(\chi + \frac{A_M(-iu)\chi A_M(iu)}{\mathcal{A}_M(i u)}\right) \, ,
\end{equation}
where the normalization factor $c_0$ is computed in Appendix~\ref{sec:appendixfermionic}, see Eq. \eqref{eq:c0def}.
In the Appendix, we prove that $\Psi_0$ satisfies the Majorana anticommutation relation $\{\Psi_0,\Psi_0\} = 2$ and that it commutes with the transfer matrices $A_M(v)$ for all $v$, and therefore is invariant under time evolution  (namely it corresponds to energy $\varepsilon_0 = 0$). Therefore,  its only contribution to the dynamics is a finite constant offset in the expectation value of $\chi$. Simlar to Ref.~\cite{vona2014exact}, we have numerically tested that the decomposition~\eqref{eq:chi-expansion} is complete, namely that no other operator except single fermionic modes appear in the expansion of $\chi$.

The expansion coefficients can be computed following \cite{vona2014exact}. They read
\begin{equation}
     c_{s\neq0} = -\frac{1}{N_s}(2 i u_s x_{M-2}x_{M-1}y_M)^2 \mathcal{A}_{M-3}(i u_s)\,,
\end{equation}
with the normalisation factor $N_s$ given in \eqref{eq:norm}.
To compute the circuit evolution of the edge operator starting from some initial (spin) state $\ket{\psi_0}$, we substitute the time evolution of the fermion modes \eqref{eq:fermion-time-evolution} into \eqref{eq:chi-expansion}
\begin{equation}
    \langle\chi(t)\rangle = \langle\psi_0|\chi(t)|\psi_0\rangle = \sum_{s = -S}^{S}c_s(t)\langle\psi_0|\Psi_s|\psi_0\rangle \, ,
\end{equation}
with
\begin{equation}
    c_s(t) = c_s \left(\frac{i u_s - 1}{i u_s +1}\right)^t\,.
\end{equation}

We are now left with two problems. First, we need to construct the polynomial $\mathcal{A}_M(u)$ and find all the roots $i u_s$. This task can be carried out efficiently, as the computational cost to find all the complex roots of a polynomial of degree $n$ scales as $O(n^3)$~\cite{aurentz2015fast}. Second, we need to compute the initial expectation values $\langle\psi_0|\Psi_s|\psi_0\rangle$. We show below that this task can also be carried out efficiently.

\begin{figure}[h]
    \centering
    \includegraphics[width=0.6\linewidth]{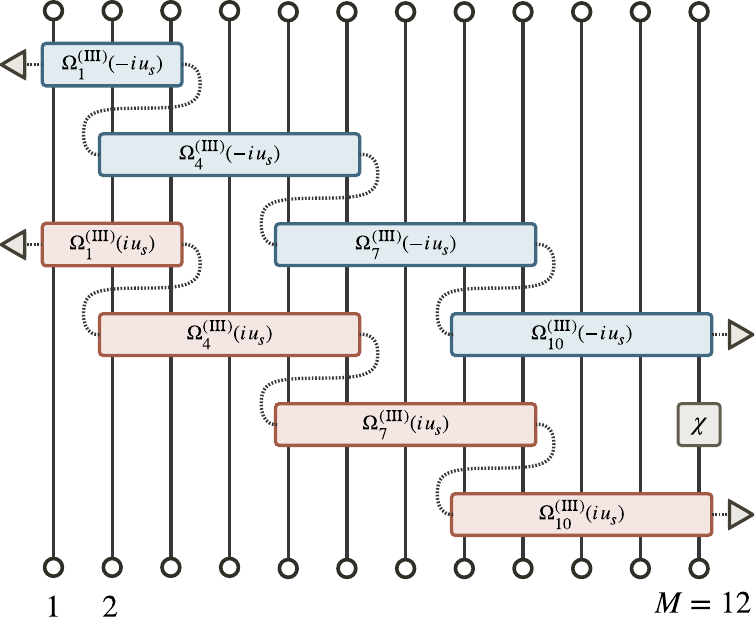}
    \caption{Tensor network computation of the expectation values of the fermionic modes in the initial state $\langle\psi_0|\Psi_s|\psi_0\rangle$ for $M = 12$. The solid lines represent the (physical) spin degrees of freedom, while the dashed legs correspond to the auxiliary space introduced in the MPO construction \eqref{eq:ABCIIIMPO}. The triangles represent the left and right vectors appearing in the same equation.}
    \label{fig:expectation-value-mpo}
\end{figure}

To this end, we use the definition of the fermionic modes \eqref{eq:fermion-def} and the MPO form of the transfer matrix \eqref{eq:ABCIIIMPO} and rewrite $\langle\psi_0|\Psi_s|\psi_0\rangle$ as a tensor network built up from the $\Omega_m^{\rm (III)}(i u_{\pm s})$ and the edge operator $\chi$ (see Fig. \ref{fig:expectation-value-mpo}). Here we focus on (arbitrary) product states, although this construction allows for arbitrary initial MPS states with non-zero initial entanglement.

Computing the initial expectation values $\langle\psi_0|\Psi_s|\psi_0\rangle$ corresponds to contracting the tensor network displayed in Fig. \ref{fig:expectation-value-mpo}, which can be done straightforwardly. Note that, although the neighboring $\Omega_m^{\rm (III)}$ tensors do not commute, leading to a staircase construction, contracting the appropriate indices only involve $2M/3$ tensors (plus the one corresponding to the edge operator). Using standard consideration from tensor-network theory~\cite{schollwock2011density}, we see that computing the initial value of every fermion mode $\Psi_s$ can be carried out at a computational cost scaling only polynomially (rather than exponentially) in the sysetm size $M$.

There is a remark in order regarding the Majorana mode $\Psi_0$ appearing in the expansion \eqref{eq:chi-expansion}. Although the above construction works for the fermionic modes with $s\neq0$, the Majorana mode is more involved. Luckily, it is simple to compute the initial expectation value $\langle\psi_0|\chi|\psi_0\rangle$ having chosen a specific representation (here we use $\chi = Z_M$). Then, the contribution of $\Psi_0$ to the initial value can be easily computed as $c_0\langle\psi_0|\Psi_0|\psi_0\rangle = \langle\psi_0|\chi|\psi_0\rangle - \sum_{s\neq0}c_s\langle\psi_0|\Psi_s|\psi_0\rangle$.

\begin{figure}[h]
    \centering
    \includegraphics[width=1\linewidth]{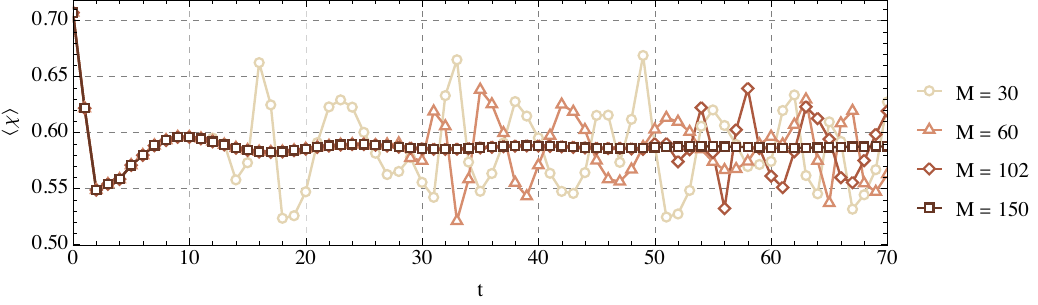}
    \caption{Time evolution of the expectation value of the edge mode under the circuit dynamics $\mathcal{V}^{\rm (III)}_M$ for multiple values of $M$ starting from the initial (spin) state $|\psi_0\rangle = \otimes_{m = 1}^{M}(\cos\vartheta|0\rangle+\sin\vartheta|1\rangle)$ with $\vartheta = \pi / 8$. The results correspond to $\phi_m = 1$.}
    \label{fig:dynamics}
\end{figure}

Having obtained the initial values of the fermionic modes and the roots $i u_s$, the dynamics is essentially solved. An example of the discrete time evolution of the edge mode $\chi$ is presented in Fig. \ref{fig:dynamics} for an initial state $|\psi_0\rangle = \otimes_{m = 1}^{M}(\cos\vartheta|0\rangle+\sin\vartheta|1\rangle)$ with $\vartheta = \pi / 8$. The plot shows data up to system sizes $\sim 150$ and time~$70$, which would be unfeasible using standard tensor-network or exact-diagonalization approaches. The phases are set to $\phi_m = 1$, corresponding to a homogeneous circuit. The results are cross-checked with exact-diagonalization of the spin-system for small system sizes, finding perfect agreement.

\section{Outlook}
\label{sec:outlook}

We have considered different families of local quantum circuits and proved that they feature FFD, similar to the model originally introduced in Ref. \cite{fendley2019free}. The building blocks of the circuits are unitary gates that are given in terms of Hamiltonian densities satisfying the FFD algebra. Such circuits were previously conjectured to be free fermionic in Ref.~\cite{fukai2025quantum}. Although some of the circuits considered here have been previously shown to be FFD solvable~\cite{fendley2019free, fukai2025free}, our construction allows us to treat the different families of circuits on equal footing by embedding the evolution operator in a family of mutually commuting transfer matrices, proving all conjectures of Ref.~\cite{fukai2025quantum}. Going further, we have addressed the possibility to classically simulate the circuits. We proved that the dynamics of certain local observables can be computed efficiently, with the computational cost related to that of simple tensor network contractions, scaling polynomially (rather than exponentially) in the system size $M$. 

Our work opens several research directions and raises new questions. First, our results motivate developing a systematic approach to express spin operators in terms of fermionic ones. We expect that progress beyond the state of the art~\cite{vona2014exact} could be made exploiting the recent results of Ref.~\cite{vernier2025hilbert}, which completed the FFD operator algebra in terms of new ancillary fermions. We believe that such ancillary fermionic operators may play a role in expressing arbitrary spin operators in terms of the fermionic degrees of freedom.

Similarly, it would be interesting to characterize the states which are mapped onto fermionic Gaussian states via the FFD mapping. This would make it possible to investigate the dynamics of the system beyond local observables. For instance, it would be especially interesting to study the dynamics of entanglement entropy. Indeed, given the free-fermionic spectrum, one may expect that an effective quasi-particle picture~\cite{calabrese2006time,fagotti2008evolution,calabrese2016quantum} could be derived, achieving an analytic description of the entanglement dynamics in the thermodynamic limit. 

Finally, another natural direction would be to investigate random Floquet circuits with FFD. Technically, this could be achieved by defining ensembles of circuits with a fixed architecture, where the parameters $\phi_m$ in Eq.~\eqref{eq:gi} are random variables, i.i.d. over the different gates forming the Floquet operator (the Floquet operator, however, does not change in time, so that the disorder is only in the space direction). It would be especially interesting to understand whether the physical properties of these ensembles (such as the late-time average bipartite entanglement entropy) can be understood in terms of standard fermionic random Gaussian ensembles~\cite{liu2018quantum,bernard2021entanglement,bianchi2021page}. We leave these directions for future work.

\section*{Acknowledgments}
We thank Paul Fendley for insightful discussions, and Bal\'azs Pozsgay, Kohei Fukai, and István Vona  for pointing out the non-locality of the logarithmic derivative of the transfer matrix of circuit III. EV acknowledges support from the CNRS-IEA. The work of DSS and LP was funded by the European Union (ERC, QUANTHEM, 101114881). Views and opinions expressed are however those of the author(s) only and do not necessarily reflect those of the European Union or the European Research Council Executive Agency. Neither the European Union nor the granting authority can be held responsible for them.

\begin{appendix} 

\section{Proof of identities for the fermion operators}
\label{sec:appendixfermionic}

In this appendix, we provide additional details on the quantum circuits I, II, and III defined in the main text. For all three circuits we define for generic $v$, 
\be 
\Psi(v) = A_M(v) \chi A_M(-v)
\ee 

\subsection{Circuit I}

Using the recursion relations \eqref{recursionAIu}, together with the fact that $A_{M-1}(u)$ (resp. $B_{M-1}(u)$) commutes (resp. anticommutes) with $\chi$, we have for all $u,v$
\begin{align}
\{A_M(u),\Psi(v)  \}
&= 2 x_M A_M(v) \chi A_{M-1}(u) A_M(v)  
\label{eq:comAPsiI}
\\
[A_M(u), \Psi(v) ]
&= -2 y_M A_M(v) \chi B_{M-1}(u) A_M(v)  
\end{align} 

In the RHS of both equations, we can use the recursion \eqref{recursionAIu} to reexpress $A_M(v)$ and the algebra \eqref{eq:ABCalgebra} to commute it with $A_{M-1}(u)$ and $B_{M-1}(u)$ respectively. Combining both equations, we get as a result 
\be 
u \{A_M(u), \Psi(v) \} - v [A_M(u), \Psi(v) ]
=
2 \mathcal{A}_M(v) (u x_M A_{M-1}(u) + v y_M B_{M-1}(u)) \chi \,.
\ee 
The above relation is true for any $v$. When $v$ is one of the roots of $\mathcal{A}_M(v)$, $v=i u_k$, the right-hand side cancels, and we get 
\be 
u \{A_M(u), \Psi(i u_k) \} = i u_k [A_M(u), \Psi(i u_k) ] \,,
\label{eq:APsiI}
\ee 
which can immediately be rearranged into Eq.~\eqref{eq:fermion-time-evolution} in the main text.
Another useful relation, true for any $v$, can be proved from \eqref{recursionAIu}: 
\be 
\{ \chi , \Psi(v) \} = 2(-\mathcal{A}_{M}(v)+ 2x_M^2 \mathcal{A}_{M-1}(v) )  \,.
\label{eq:comchiPsi}
\ee 
We can use this to compute the anticommutators of fermionic modes: when $v=i u_k$ is some root of $\mathcal{A}_M$, we have:
\begin{align} 
\{\Psi(i u_k) , \Psi(u) \} &= \frac{i u_k -u}{i u_k +u} {A}_M(u) \{\Psi(i u_k), \chi\} {A}_M(-u)
\\&=4 \frac{i u_k -u}{i u_k +u}  x_M^2 \mathcal{A}_{M-1}(i u_k)  \mathcal{A}_M(u)  \,.
\end{align}
where in the first line we have used \eqref{eq:APsiI}, and in the second line we have used \eqref{eq:comchiPsi}.
This relation is true for any $u$, however we can now specialize $u$ to some root of $\mathcal{A}_M$ : if $u=iu_l \neq - i u_k$, the right-hand side simply vanishes. 
If $u=-i u_k$, the vanishing of $\mathcal{A}_M(-i u_k)$ compensates for that of the denominator $i u_k+u$, and we get the expression of the anticommutator from a limiting procedure: 
\be 
\{\Psi(i u_k), \Psi(-i u_k)\} 
= 8 i u_k x_M^2 \mathcal{A}_{M-1}(i u_k)  \mathcal{A}_M'(-i u_k) \,,
\ee 
which defines the normalization factors $N_k$ given in the main text, eq. \eqref{eq:norm}.


\subsection{Circuit II}

Analogously to the case of circuit I, we derive, using the recursion formula \eqref{recursionAII} and the algebra \eqref{eq:ABCalgebra} the following relation, valid for any $u,v$: 
\be 
u \{A_M(u), \Psi(v) \} - v [A_M(u), \Psi(v) ]
= 2  \mathcal{A}_M(v) \left( v A_{M}(u)+ (u-v) x_{M-1} x_M A_{M-2}(u)   \right) \chi \,,
\label{APsiII}
\ee 
where in the expansion \eqref{recursionAII} we have used that some terms commute (resp. anticommute) with $\chi$ (we recall that for circuit II the latter obeys a slightly different definition from that of circuits I and III, see eq. \eqref{eq:chiII}).
 When $v$ is one of the roots of $\mathcal{A}_M(v)$, $v=i u_k$, the right-hand side of eq. \eqref{APsiII} cancels, and we get, as for circuit I, the relation \eqref{eq:APsiI}, which leads to eq.~\eqref{eq:fermion-time-evolution} in the main text.

We also get from \eqref{recursionAII} the following formula, true for any $v$: 
\be 
\{ \chi , \Psi(v) \}   
= -2 \mathcal{A}_{M}(v)  +  (2  x_{M} x_{M-1})^2  \mathcal{A}_{M-2}(v)
\,.
\label{eq:comchiPsiII}
\ee 
As for circuit I we can use this to compute the anticommutators of fermionic modes: when $v=i u_k$ is some root of $\mathcal{A}_M$, we have:
\begin{align} 
\{\Psi(i u_k) , \Psi(u) \} &= \frac{i u_k -u}{i u_k +u} A_M(u) \{\Psi(i u_k), \chi\} {A}_M(-u)
\\&=\frac{i u_k -u}{i u_k +u}   (2 i u_k x_{M-1} x_{M})^2  \mathcal{A}_{M-2}(i u_k)  \mathcal{A}_M(u)  \,.
\end{align}
where in the first line we have used \eqref{eq:APsiI}, and in the second line we have used \eqref{eq:comchiPsiII}.
Again, we can use this to prove the canonical anticommutation relations, and determine the normalization factor. Setting $u=-i u_k$ in the above,  we find
\be 
\{\Psi(i u_k), \Psi(-i u_k)\} 
= 8 i u_k   ( x_{M-1} x_{M})^2  \mathcal{A}_{M-2}(i u_k)
\mathcal{A}_M'(-i u_k) \,,
\ee 
which defines the normalization factors $N_k$ given in the main text, Eq.~\eqref{eq:norm}.

\subsection{Circuit III}

Analogously to the other cases, we derive, using the recursion formula \eqref{recursionAIII} and the algebra \eqref{eq:ABCalgebra} the following relation, valid for any $u,v$: 
\be 
u \{A_M(u), \Psi(v) \} - v [A_M(u), \Psi(v) ]
= 2 u \mathcal{A}_M(v) \left(A_M(u) + i (v-u)x_{M-2}x_{M-1}y_{M} A_{M-3}(u) h_M \right) \chi \,.
\ee 
 When $v$ is one of the roots of $\mathcal{A}_M(v)$, $v=i u_k$, the right-hand side cancels, and we get, as for circuits  I and II, the relation \eqref{eq:APsiI}.

We also get from \eqref{recursionAIII} the following formula, true for any $v$: 
\be 
\{ \chi , \Psi(v) \} = 2 \mathcal{A}_{M}(v)  -  (2 v x_{M-2} x_{M-1}y_M)^2  \mathcal{A}_{M-3}(v)  \,.
\label{eq:comchiPsiIII}
\ee 
Proceeding as in circuit I and circuit II, we have that when $v=i u_k$ is some root of $\mathcal{A}_M$,
\begin{align} 
\{\Psi(i u_k) , \Psi(u) \} &= \frac{i u_k -u}{i u_k +u} A_M(u) \{\Psi(i u_k), \chi\} {A}_M(-u)
\\&=-\frac{i u_k -u}{i u_k +u}  (2 i u_k x_{M-2} x_{M-1}y_M)^2  \mathcal{A}_{M-3}(i u_k) \mathcal{A}_M(u)  \,,
\end{align}
where in the first line we have used \eqref{eq:APsiI}, and in the second line we have used \eqref{eq:comchiPsiIII}.
The normalization factor $N_k$ follows, as in the other cases, by setting $u=-i u_k$, 
\be 
\{\Psi(i u_k), \Psi(-i u_k)\} 
= -8 i u_k  (i u_k x_{M-2} x_{M-1}y_M)^2  \mathcal{A}_{M-3}(i u_k) 
\mathcal{A}_M'(-i u_k) \,.
\ee

Lastly, we prove that the zero mode $\Psi_0$ defined as in Eq.~\eqref{eq:psi_0_def} indeed squares to identity and commutes with the transfer matrices $A_M(v)$ (and therefore with the evolution operator $\mathcal{V}_M^{\rm (III)}=A_M(1)$). 
Let us start by defining  
\be 
\mathcal{Q}(i u) =  \frac{1}{2}\left(\chi + \frac{A_M(iu) \chi A_M(-i u)}{\mathcal{A}_M(iu)} \right)  \,.
\ee 
Using Eq. \eqref{eq:comchiPsiIII}, we find 
\be 
\mathcal{Q}(i u)^2 =  1 + \frac{ (u x_{M-2} x_{M-1} y_M)^2 \mathcal{A}_{M-3}(i u)}{\mathcal{A}_M(u)} \,. 
\label{Qsquared}
\ee 
The polynomials $\mathcal{A}_M(u)$ and $\mathcal{A}_{M-3}$ have respective degrees $S$ and $S-1$ in $u^2$. Therefore, \eqref{Qsquared} has a finite limit as $u\to\infty$. Defining $c_0$ as 
\be 
\label{eq:c0def}
c_0^2 = 1 + \lim_{u\to \infty} \frac{ (u x_{M-2} x_{M-1} y_M)^2 \mathcal{A}_{M-3}(i u)}{\mathcal{A}_M(u)}  \,,
\ee
we indeed have that zero mode $\Psi_0$ defined as \eqref{eq:psi_0_def} squares to the identity.
We next show that it commutes with the transfer matrices $A_M(v)$, for any $v$.
A simple use of the recursion relations \eqref{recursionAIII} and algebra \eqref{eq:ABCalgebra} yields 
\be
\frac{
[A_M(v) , \mathcal{Q}(iu)]}{i v x_{M-2} x_{M-1} y_M } 
 =  \frac{A_M(i u) \left(
 2 x_{M-1}x_{M-2} x_M A_{M-3}(-iu) A_{M-3}(v) + [A_M(v),A_{M-3}(-iu)]
 \right) h_M \chi}{\mathcal{A}_M(iu)} \,. 
\ee 
By construction, $A_M(iu)$ is of degree $S$ in $u$, and $A_{M-3}(iu)$ is of degree $S-1$. Therefore, the right-hand-side of the above equation vanishes in the $u\to \infty$ limit. We therefore have 
\be 
\lim_{u\to \infty}
\left[ \mathcal{Q}(iu) ,A_M(v)\right] 
=0 \,, 
\ee 
which proves the commutation $[\Psi_0,A_M(v)]=0$.

\end{appendix}

\bibliography{bibliography}
	
\end{document}